# Re: Re: A contribution to the history of quarks …

Fyodor V. Tkachov

Institute for Nuclear Research of Russian Academy of Sciences
Moscow 117312
Russian Federation

*Abstract.* The emergence of Boris Struminsky's January, 1965 paper with a footnote that introduced a new quark quantum number now known as color caused a response [arXiv:0908.2772] that is seen, perhaps contrary to what it was intended to convey, to corroborate the general picture that comes out of the evidence summarized in [arXiv:0904.0343].

The emergence of Boris Struminsky's paper [1] in ref. [2] caused a response [3] that rather surprised me by a blatant manner of construing the findings and testimonies of [2]. I am forced to post these comments to prevent a situation where ref. [3] could be pointed to as refuting "One False Revision" (here and below I quote terms and phrases from v1 of [3] dated 19 Aug 2009).

Ref. [3] starts out with a picture of how it "was generally believed that the Soviet contribution was made in" a paper by Bogoliubov, Struminsky and Tavkhelidze. Then it is stated that this "opinion" (Mr. Petrov's own word) "was contested in very strong terms" in [2]. Well, opinions are opinions, and facts are facts. If the fact of existence of Struminsky's solo paper somehow shatters the "opinion", so much the worse for the "opinion". Aren't we physicists.

On page 2, Mr. Petrov proposes that "the author of" [2] came "to an unambiguous conclusion" which Mr. Petrov — as if lacking confidence in the power of his arguments — sets in bold italic. The "conclusion" is said to be that "The Soviet contribution into the discovery of color is due to solely B.V. Struminsky." Then Mr. Petrov quotes the passages from ref. [2] that he claims somehow "express" that conclusion, acknowledging them to have been formulated "quite carefully". The truth is, the passages accurately state simple facts about the content of [1], and the honor of inventing the "conclusion" is entirely Mr. Petrov's.

After that, the deconstruction of ref. [2] reaches its climax in an assertion that "the author of" [2] brought the reader to an "implicit but factual" (whatever that means) "conclusion". The "conclusion" which Mr. Petrov draws from ref. [2] — and sets (again) in bold italic — is that "N.N. Bogoliubov committed an immoral deed: he «hanged on» Struminsky's discovery". Mr. Petrov did not explain exactly how he came to such a conclusion because other people (at least those who I discussed the matter with) tend to draw a different conclusion from the evidence of ref. [2], namely, that Bogolyubov rather deserves credit for allowing his PhD student to publish ref. [1] solo. Whereas the charge of immorality should more aptly be directed at those who have been trying, in a concerted effort over a number of years after Struminsky's and Bogoliubov's demise, to obliterate Struminsky's name and replace it with another one in the story of how the "Soviet contribution" to the discovery of quark color was made.

(Speaking of obliterations. Calling names ("home-bred Sherlock Holmes") is a poor explanation for how all the library catalogue cards for Struminsky disappeared from the front-desk catalogue in the JINR Library — a fact that was independently verified at my request. Mr. Petrov should try again.)

It follows that the "grave and blasphemous invectives" mentioned at the top of page 3 of [3] are only Mr. Petrov's contrivance.

In the list of "considerations" on pages 3 and 4 Mr. Petrov presents evidence for how much respect Struminsky had for his scientific advisor, how none of Bogoliubov's "disciples and co-workers" doubted his decency, etc. There is no particular reason to doubt all that, even if one should not take for granted anything anyone (especially persons in subordinate positions) publicly said about a powerful Soviet Academician and Director. Along the way, Mr. Petrov records his mental struggle with his own imagination, but who cares.

Right after the list Mr. Petrov points out a "clear way out of this false problem". The "problem" seems to be the contradiction between the "conclusion" invented by Mr. Petrov and a general opinion of what a nice guy Nikolai Bogoliubov was. I agree that the "problem" is "false" in the sense that there is no contradiction between that opinion and the evidence collected in [2].

Anyhow, the "clear way" is, according to Mr. Petrov, "to acknowledge that Bogoliubov has informed his PhD student Struminsky in general terms on resolution of the quark statistics problem". Again, Mr. Petrov chooses one interpretation where at least one alternative exists, namely, that the "resolution" may have emerged from a lively discussion between two researchers attacking a scientific enigma, as it sometimes happens (cf. the two photos in the Appendix). For fairness' sake, one cannot exclude a possibility that Mr. Petrov has never had such an experience, which could excuse his oversight.

But let us assume this hypothesis to be true as Mr. Petrov insists it is. What then? Then, it simply follows that the "Soviet contribution" to the discovery of quark color is strictly a matter between Bogoliubov and Struminsky. For Mr. Petrov will surely agree — as all "Bogoliubov's disciples and co-workers" not doubt should — that Nikolai Bogoliubov, whose decency it would be a blasphemy to doubt, could not have possibly committed an immoral deed of excluding a third party from that scientific interaction if the party was worthy of inclusion. This very point was emerging from the evidence collected in ref. [2], and it now receives a corroboration all the more capital that it has come from a most unlikely source.

---

An intelligent reader would be puzzled by the cheap cheat attempted in ref. [3]. The following quite well-known facts — easily verifiable via Web search, at least within its Russian segment — might be the missing links to help one find a plausible explanation: Mr. Petrov has been a *combatant* of Academician A.A. Logunov, formerly an all-powerful Director and still a powerful Scientific Advisor of the Institute of High Energy Physics in Protvino where Mr. Petrov is head of the same theory department of which Mr. Tavkhelidze served head in the late 1960's. Academician Logunov also happens to be a collaborator of Mr. Tavkhelidze, their relationship going back at least to their residence in Dubna in the early 1960's.

**Appendix**

On the two photos below Boris Struminsky is on the left, Nikolai Bogoliubov is on the right. A rapport between the two is plainly seen.

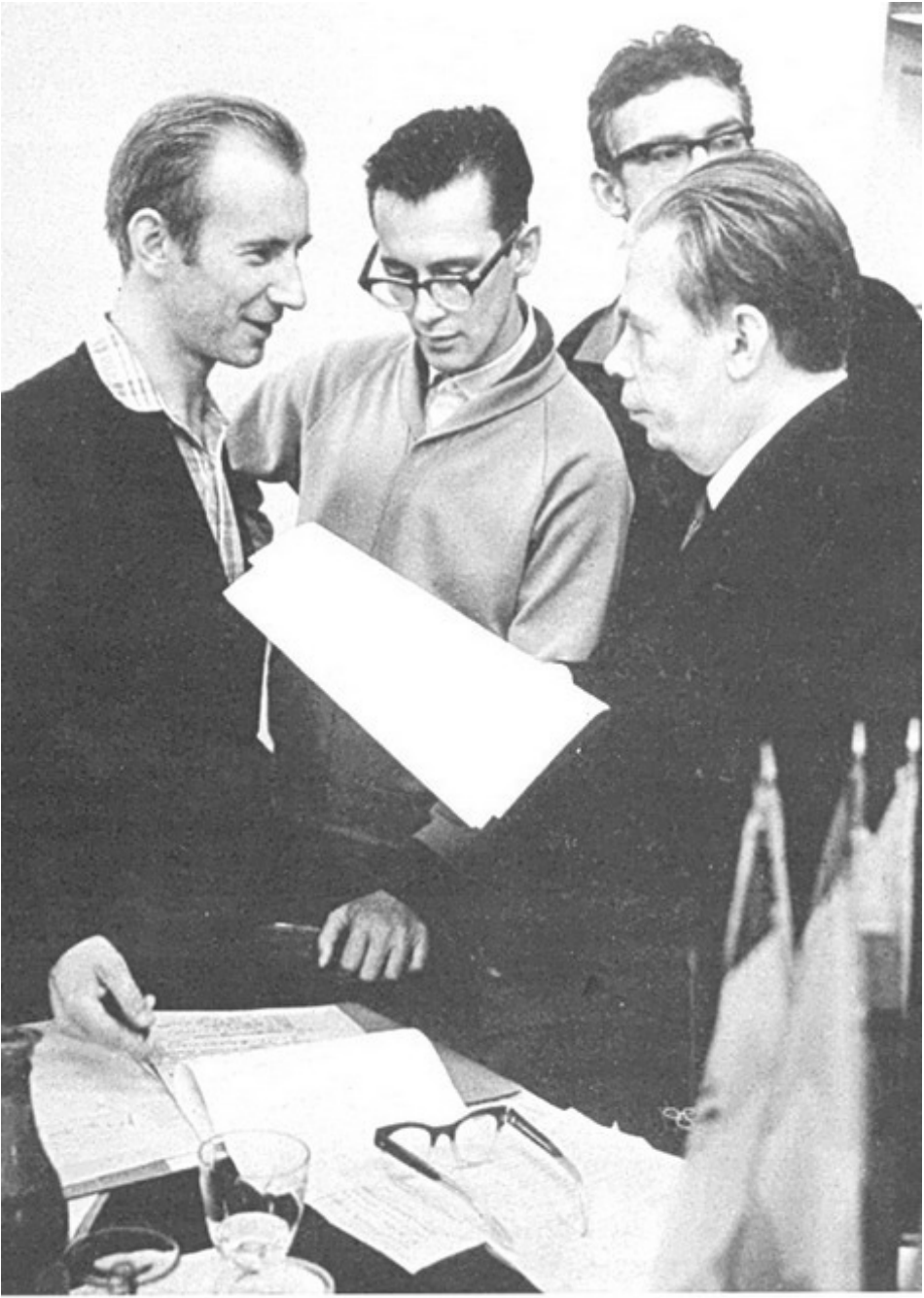

Б.В.Струминский, В.А.Матвеев, Н.Н.Боголюбов. Дубна, 1969

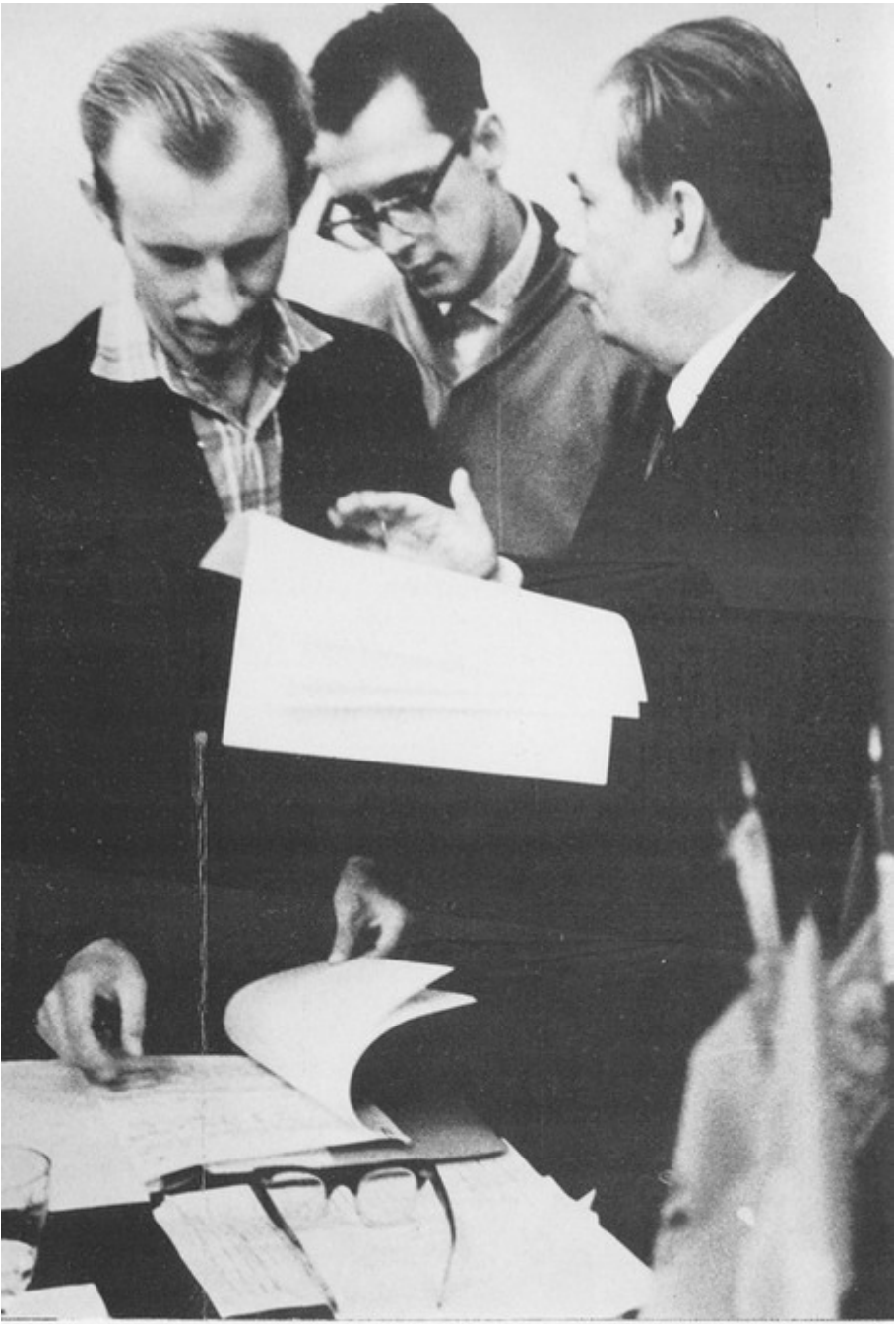

Б.В.Струминский, В.А.Матвеев, Н.Н.Боголюбов. Дубна, 1969 г.

The very young man in between is Viktor Matveev, currently Academician and a Director, and the one who took public responsibility for nominating Mr. Tavkhelidze for the 2009 Bogoliubov Award for the discovery of quarks' color, at a session of the Scientific Council of INR RAS in February, 2009.

More photos and comments can be found at
http://www.inr.ac.ru/~ftkachov/struminsky/photos_strum_bog.htm